\documentclass[twocolumn,showpacs,prl]{revtex4}
\usepackage{amsmath}
\usepackage{epsf}
\usepackage{graphicx}
\usepackage{dcolumn}
\usepackage{bm}

\begin{document}

\title{Universality in scattering by large-scale potential fluctuations\\
in two-dimensional conductors}
\author{Vladimir I. Yudson$^{1}$ and Dmitrii L. Maslov$^{2}$}
\date{\today}

\begin{abstract}
We study electron propagation through a random array of rare,
opaque and large (compared the de Broglie wavelength of electrons)
scatterers. It is shown that for any convex scatterer the ratio of
the transport to quantum lifetimes $\eta=\tau_{\rm tr}/\tau_{{\rm
tot}}$ does not depend on the shape of the scatterer but only on
whether scattering is specular or diffuse and on the spatial
dimensionality ($D$). In particular, for specular scattering, $\eta$ is
a universal constant determined only by the dimensionality 
of the system: $\eta = 2$ for $D = 3$ and $\eta = 3/2$ for $D =
2$. The crossover between classical and quantum regimes of
scattering is discussed.
\end{abstract}

\pacs{72.10-d.,72.10.Fk,72.15.Lh}

\affiliation{ $^1$Institute of
Spectroscopy, Russian Academy of Sciences, Troitsk, Moscow region,
142190, Russia\\$^{2}$Department of Physics, University of
Florida, P. O. Box 118440, Gainesville, FL
32611-8440}

\maketitle

At low temperatures, electron transport is controlled by disorder.
The conventional model for disorder is an ensemble of relatively
sparse impurities or surface roughness, whereas larger
imperfections, greatly exceeding the electron wavelength, are
assumed to be prevented by a reasonably advanced growth technique.
This model is facing certain difficulties in view of mounting
evidence for macroscopic inhomogeneities in a number of key
materials, \emph{e.g.}, high-T$_{c}$ superconductors, manganites,
semiconductor heterostructures, etc. For example, nanoscale inhomogeneities in high-T$%
_{c}$ materials and manganites have been observed via scanning
tunnelling microscopy \cite{cuprates,mang}. Also, the ionized
impurities in very high-mobility GaAs heterostructures are
engineered to be so far away from the 2D gas layer plane that it
is not these impurities but rather large-scale potential
fluctuations that provide the dominant scattering mechanism for
electrons. Sometimes the large-scale potential fluctuations are
introduced intentionally, as antidot arrays \cite{antidot}.
Randomness in positions and/or shapes of antidots leads to
additional scattering of electrons. On the theoretical side, the
semiclassical motion of electrons in the presence of large-scale
inhomogeneities has attracted a significant interest due to
non-Boltzmann effects in magneto- and ac transport
\cite{karlsruhe}. In this communication, we study some very basic
yet, to the best of our knowledge, unexplored universalities in
the effective scattering cross-sections by large-scale
fluctuations.

\begin{figure}[tbp]
\begin{center}
\epsfxsize=1.0\columnwidth\epsffile{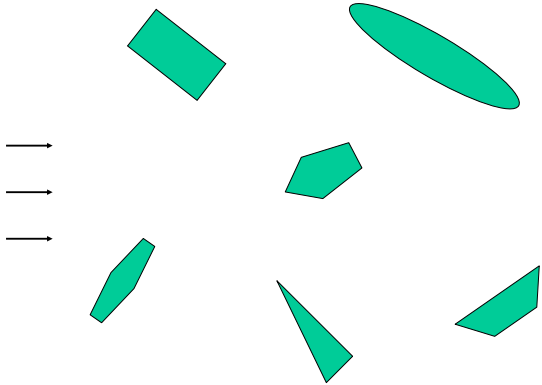}
\end{center}
\caption{Model of disorder.}
\label{fig:fig1}
\end{figure}

We assume that disorder is represented by an ensemble of large (of size $%
a$ much larger than the electron de Broglie wavelength $\lambda $)
objects of irregular but smooth shape, placed and oriented
randomly along the conducting plane, see Fig.~\ref{fig:fig1}. An
important parameter characterizing the spatial structure of
disorder is the ratio of transport and ``quantum'' mean free
times,
\begin{equation}
\eta =\frac{\tau _{\mathrm{tr}}}{\tau _{\mathrm{q}}}=\frac{\sigma _{\mathrm{%
tot}}}{\sigma _{\mathrm{tr}}},  \label{alpha}
\end{equation}
associated with the transport
\begin{equation}
\sigma _{\mathrm{tr}}=\int d\Omega \frac{d\sigma }{d\Omega }\left( 1-\cos
\theta \right)  \label{sigmtr}
\end{equation}
and total
\begin{equation}
\sigma _{\mathrm{tot}}=\int d\Omega \frac{d\sigma }{d\Omega }
\label{sigmatot}
\end{equation}
scattering cross-sections, correspondingly.

Experimentally, $\tau _{\mathrm{%
tr}}$ is extracted from the conductivity whereas $\tau _{{\mathrm{q}}}$ is
obtained as a damping rate of de Haas-van Alfen or Shubnikov-de Haas
oscillations. For long-range disorder, $\eta \gg 1$. A classic example of
such a system is a GaAs heterostructure with modulation doping \cite{ando}.
For isotropic impurities, $\eta =1$. This is the case, \emph{e.g.} for
Si-based field-effect transistors in a certain range of densities \cite{ando}%
. $\eta <1$ corresponds to enhanced backscattering. A minimum value of $\eta
=1/2$ is achieved for the limiting case of strict backscattering, when $%
d\sigma /d\Omega \propto \delta (\theta -\pi )$.

In the present paper, we show that for a wide class of randomly placed and
oriented plane scatterers of mesoscopic size ($a\gg \lambda $), there is a
surprising universality in the value of the parameter $\eta $. Namely,
\begin{equation}
\eta =\frac{3}{2}\,  \label{eta-spec}
\end{equation}
for the case of specular scattering and
\begin{equation}
\eta =\frac{4}{3}\,\label{eta-diff}
\end{equation}
for the case of diffuse scattering. The universality of $\eta $ results from
a combination of universal features of scattering at the true quantum or
even classical levels of consideration. Given the mesoscopic size of the
objects and smoothness of their shapes, we adopt first the classical
mechanics to describe electron scattering. Later on, we discuss and
implement the important modifications caused by the quantum nature of
scattering particles.

\begin{figure}[tbp]
\begin{center}
\epsfxsize=1.0\columnwidth\epsffile{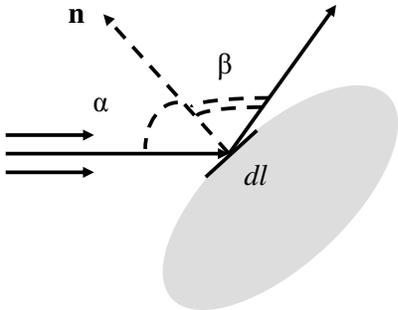}
\end{center}
\caption{Scattering by a convex object.}
\label{fig:fig2}
\end{figure}

For classical specular scattering by a sphere, $\sigma _{\mathrm{tot}%
}^{{\rm cl}}=\sigma _{\mathrm{tr}}^{{\rm cl}}$, i.e., $\eta ^{{\rm cl}}=1$ (here and
thereafter, the superscript $``cl"$ denotes classical quantities). For the
2D case of interest, the classical differential cross-section for specular scattering
by a disk of radius $a$ is given by:
\begin{equation}
\frac{d\sigma ^{{\rm cl}}}{d\Omega }=\frac{d\sigma ^{{\rm cl}}(\theta )}{d\theta }=\frac{%
a}{2}\sin {\frac{\theta }{2}},\,\,\,\,\,\,\,\,\theta \in (0,2\pi )\,,
\label{disk-diff}
\end{equation}
so that
\begin{equation}
\sigma _{\mathrm{tot}}^{{\rm cl}}=2a\,,\,\,\,\sigma _{\mathrm{tr}}^{{\rm cl}}=8a/3\,,\,\,%
\mathrm{and}\,\,\,\eta ^{{\rm cl}}=3/4\,.  \label{disk}
\end{equation}
Somewhat unexpectedly, a 2D array of randomly oriented ellipses,
squares and even rectangles is characterized by the same value of
$\eta ^{{\rm cl}}$. Here we will show that such a universality is
not accidental but is a general feature for scattering off a wide
class of (randomly oriented) plane objects. Namely, this is a
property of all convex objects. To prove this statement, consider
scattering of a particle beam (parallel to the $x$-axis) by a
small element (of length $dl$) of the boundary (see Fig.
\ref{fig:fig2}). For a general case of partially diffusive
scattering, the outgoing angle $\beta $ differs from the angle of
incidence $\alpha $, the latter is determined as the angle
between the external normal $\mathbf{n}$ to the element $dl$ and the $x$%
-axis. The distribution probability of angles $\beta \in (-\pi /2,\pi /2)$
is characterized by a function $P(\beta ;\alpha )$. In particular cases of
specular and absolutely diffuse scattering reflection $P(\beta ;\alpha
)=\delta (\beta -\alpha )$ and $P(\beta ;\alpha )=1/\pi $, correspondingly.
The scattering angle $\theta $ is given by $\theta =\pi -\alpha -\beta $ and
the flux of particles colliding with the element $dl$ is proportional to $%
\cos {\alpha }$. Assuming random orientations of scatterers in the
conduction plane, we average the cross-section over all possible
orientations of $\mathbf{n}$, i.e. over the values of $\alpha \in (-\pi ,\pi
)$. Taking into account that scattering by element $dl$ occurs only for $%
\alpha \in (-\pi /2,\pi /2)$, we obtain its contribution to the averaged
differential cross-section as
\begin{eqnarray}
\frac{d\sigma ^{{\rm cl}}(\theta )}{d\theta } &=&dl\int_{-\pi /2}^{\pi /2}\frac{%
d\alpha }{2\pi }\cos {\alpha }  \notag  \label{dsigma-gen} \\
&&\times\int_{-\pi /2}^{\pi /2}d\beta \,P(\beta ;\alpha )\delta (\theta -\pi
+\alpha +\beta )\,.
\end{eqnarray}
Assuming the distribution function $P(\beta ;\alpha )$ to be uniform along
the scatterer boundary and integrating over $dl$, we replace $dl$ in Eq.(\ref
{dsigma-gen}) by the perimeter of the object, $P$. Therefore, the total
and transport cross-sections are given by
\begin{equation}
\sigma _{\mathrm{tot}}^{{\rm cl}}=P\int_{-\pi /2}^{\pi /2}\frac{d\alpha d\beta }{%
2\pi }P(\beta ;\alpha )\cos {\alpha }\,
\end{equation}
and
\begin{equation}
\sigma _{\mathrm{tr}}^{{\rm cl}}=P\int_{-\pi /2}^{\pi /2}\frac{d\alpha d\beta }{%
2\pi }P(\beta ;\alpha )[1+\cos {(\alpha +\beta )]}\cos {\alpha }\,.
\label{sigma-tr-gen}\end{equation}
The ratio of the two cross-sections, $\eta ^{{\rm cl}},$ is determined entirely by
function $P(\beta ;\alpha )$ and is independent of a particular geometry of
the scattering object.

Consider two particular situations of special interest. For specular
scattering, Eqs.(\ref{dsigma-gen}-\ref{sigma-tr-gen}) reduce to
\begin{equation}
\frac{d\sigma ^{{\rm cl}}(\theta )}{d\theta }=\frac{P}{4\pi }\sin {\frac{\theta }{2%
}}\,,
\end{equation}
and
\begin{equation}
\sigma _{\mathrm{tot}}^{{\rm cl}}=P/\pi \,,\,\,\,\sigma _{\mathrm{tr}%
}^{{\rm cl}}=4P/3\pi \,,\,\,\mathrm{and}\,\,\,\eta =3/4\,.  \label{spec}
\end{equation}
For a disk, these expressions coincide with Eqs.(\ref{disk-diff}) and (\ref
{disk}), respectively. For absolutely diffuse scattering, we arrive at
\begin{equation}
\frac{d\sigma ^{{\rm cl}}(\theta )}{d\theta }=\frac{P}{\pi ^{2}}\sin ^{2}{\frac{%
\theta }{2}}\,,
\end{equation}
so that
\begin{equation}
\sigma _{\mathrm{tot}}^{{\rm cl}}=P/\pi \,,\,\,\,\sigma _{\mathrm{tr}%
}^{{\rm cl}}=3P/2\pi \,,\,\,\mathrm{and}\,\,\,\eta ^{{\rm cl}}=2/3\,.  \label{diff}
\end{equation}
The universality emerges as a result of the averaging over random
orientation of scatterers and is the property of any scatterer of a convex
geometry. The convexity condition prevents repeated scattering. To
illustrate the importance of this requirement, we present a simple example
of scatterers where the repeated scattering is not excluded. This is
specular scattering by randomly oriented ``right angles'' of size $a$. We
have for averaged total and transport cross-sections:
\begin{eqnarray}
\sigma _{\mathrm{tot}}^{{\rm cl}} &=&a(2+\sqrt{2})/\pi \,,\,\,\,\sigma _{\mathrm{tr%
}}^{{\rm cl}}=a(4+\sqrt{2})/\pi \,,\,\,  \label{angle} \\
\eta ^{{\rm cl}} &=&(3+\sqrt{2})/7\approx 0.63\,.
\end{eqnarray}
These values differ considerably from those for scatterers of a convex form.

The consideration that has lead to the above universal results is
not restricted to a specific dimensionality. In 3D, averaging over
random orientations of a convex scatterer is equivalent to
averaging of a contribution of a particular surface patch which is
rotated over the whole solid angle. For instance, for specular
scattering in 3D we arrive at the
universal expressions: $\sigma _{\mathrm{tot}}^{{\rm cl}}=\sigma _{\mathrm{tr}%
}^{{\rm cl}}=S/4$ ($S$ is a surface area), and $\eta ^{{\rm cl}}=1$. In fact, the
universal connection $\sigma _{\mathrm{geom}}=S/4$ between the geometrical
cross-section and the surface area of randomly oriented convex scatterers is
well known in the field of light scattering by dust particles (see e.g.,
Ref. \cite{Light}), and it is difficult to refer to the very first proof
of this theorem.

Now we should account for modifications caused by the wave nature
of scattering particles (in optical language, we are going beyond
the geometrical optics). As long as  $\lambda /a$ is much smaller
than one, these modifications are negligible for the transport
cross-section, but are very substantial for the total
cross-section. For instance, the total cross-section of a sphere
(of
radius $a\gg \lambda $ ) is \emph{twice larger than the classical value} $%
\sigma _{\mathrm{tot}}^{{\rm cl}}=\pi a^{2}$ determined simply by the geometrical
cross-section (see, e.g., \cite{Sphere,Light}). Such a dramatic discrepancy
with the classical result (the so-called Extinction Paradox) stems from
a sharp peak in the differential cross-section for forward scattering ($%
\theta \rightarrow 0$) which cannot be described semiclassically
\cite{Sphere}. On the other hand, this peak does not contribute to
the transport cross-section due to the factor $1-\cos {\theta }$
($\rightarrow 0$, at $\theta \rightarrow 0$) and $\sigma
_{\mathrm{tr}}\approx \sigma _{\mathrm{tr}}^{{\rm cl}}$.

It is important that the relations between the classical and
quantum cross-sections, described in the previous paragraph, are
not specific either to spherically symmetric scatterers or to a
particular spatial dimensionality. To make evident the
universality of these relations, we recall the underlying physical
principle which governs scattering from an opaque object of size
$a\gg \lambda $. Directly behind this object
there is a shadow region with a vanishing amplitude of the wave field, $A=0$%
. According to the superposition principle,  $A=A_{i}+A_{s}$,
where $A_{i}$ and  $A_s$ are the amplitudes of the incident and
scattered waves, correspondingly. Therefore, $A_{s}=-A_{i}$ within
the shadow region (Babinet principle, 1837). This means, that in
addition to the flux, scattered in the directions outside the
shadow region, an opaque object also scatters the incoming
radiation in the forward direction, within a very narrow
diffraction cone of angle $\theta \sim \lambda /a$. Obviously, the
flux of the forward-scattered wave equals to the
flux of the incident wave through the geometric cross-section $\sigma _{%
\mathrm{geom}}$ of the scatterer. This leads to an additional contribution $%
\delta \sigma _{\mathrm{tot}}=\sigma _{\mathrm{geom}}$ to the
total scattering cross-section. As a result, the true total
scattering cross-section $\sigma _{\mathrm{tot}}$ is a sum of its
semi-classical value $\sigma _{\mathrm{tot}}^{{\rm cl}}=\sigma
_{\mathrm{geom}}$
and the forward scattering part $\delta \sigma _{\mathrm{tot}}=\sigma _{%
\mathrm{geom}}$. Thus, we arrive at the universal relationships
\begin{eqnarray}
\sigma _{\mathrm{tot}} &=&2\sigma _{\mathrm{tot}}^{{\rm cl}}\,,  \label{cl-q-tot}
\\
\sigma _{\mathrm{tr}} &=&\sigma _{\mathrm{tr}}^{{\rm cl}}\,  \label{cl-q-tr}
\end{eqnarray}
valid to the leading order in the small parameter $\lambda /a$ for
an arbitrary opaque scatterer with a well-defined boundary (we
will discuss the latter condition below). These relations refer to
an arbitrary orientation of the scatterer, hence they sustain
averaging over orientations. It needs to be emphasized that
Eq.(\ref{cl-q-tot}) refers to the total cross-section measured at
distances larger then the Fraunhofer length, $L_{\rm
F}=a^2/\lambda$, from the scatterer, where quantum small-angle
scattering smears the classical shadow.

Finally, combining Eqs. (\ref{cl-q-tot}) and (\ref{cl-q-tr}) with Eqs.(\ref
{spec}) and (\ref{diff}), we obtain the announced results, $\eta = 3/2$ [Eq.(%
\ref{eta-spec})] and $\eta = 4/3$ [Eq.(\ref{eta-diff})], respectively.

Although we used a two-step (classical-to-quantum) derivation of Eqs.(\ref
{eta-spec}) and (\ref{eta-diff}), it should be emphasized that their
validity range is wider than that of the intermediate expressions, which
involve the classical total cross-section $\sigma _{\mathrm{tot}}^{{\rm cl}}$. The
point is that the latter quantity is well-defined only for scatterers with a
sharp boundary. If, on the contrary, the scattering potential falls off
continuously with the distance, the classical total cross-section $\sigma _{%
\mathrm{tot}}^{{\rm cl}}$ is infinite, no matter how small the potential is away
from the center \cite{Sphere}.  This makes the very notion of the classical
total cross-section very restricted. On the contrary, the true quantum total
cross-section accounts for the weakness of scattering by the potential tail
and remains finite if the potential decays sufficiently fast \cite{Sphere}.
For instance, if the scattering potential consists of a hard core of size $%
a\gg \lambda $ and a tail falling off as  $\exp {(-(r-a)/h)}$ for $r\geq a$
with $h\ll a$, the quantum total and transport cross-sections differ from
their values for the bare core only by small corrections of the relative
order $h/a\ll 1$. This validates the robustness of the obtained universal
relations with respect to smearing of the scatterer's edge.
\begin{figure}[tbp]
\begin{center}
\epsfxsize=1.0\columnwidth\epsffile{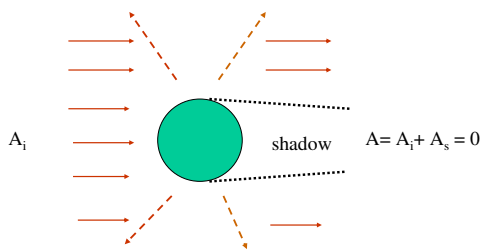}
\end{center}
\caption{Formation of a shadow by an opaque object.}
\label{fig:fig3}
\end{figure}
Experimentally, the transport scattering rate $1/\tau _{\mathrm{tr}}$ is
extracted from the conductivity.
 The
quantum decay rate, $1/\tau _{\mathrm{q}}$, may be obtained, in principle,
by measuring attenuation of an electron beam propagating through a
disordered stripe. However, it is more practical to extract $1/\tau _{%
\mathrm{q}}$ from the amplitude of the de Haas-van Alfen or  Shubnikov-de Haas
oscillations in relatively weak magnetic fields (to avoid significant
changes of the decay rate caused by the field itself). In both types of
experiments, attenuation of the measured quantities (the beam amplitude or
the amplitude of magneto-oscillations) is due to deflecting particles from their
original trajectories by scattering, no
matter
 in what direction.
Consequently, $1/\tau_q$ is related to the total cross-section. If
magneto-oscillations are measured in a very weak magnetic field,
so that the cyclotron radius is larger than any other lengthscale
of the problem (but still smaller than the system size), this
cross-section is given by the quantum expression. The value of
$\eta$, appropriate for this situation ($\eta=3/2$) reflects a
two-fold enhancement of the quantum cross-section compared to the
classical one. The classical value of $\eta^{{\rm cl}}=3/4$, which
is sometimes cited in the literature on chaos \cite{chaos} and
classical memory effects \cite{polyakov} in a system of disk
scatterers, does not correspond to the ratio of the observable
mean free times, as $\tau^{{\rm cl}}_{{\rm tot}}$ is not
observable under such experimental conditions. However, there is
an interesting but still open question about the
quantum-to-classical crossover in the effective total
cross-section with an increase of the magnetic field (i.e.,
shortening of cyclotron orbits), in analogy with the crossover for
the scattering by a single obstacle, observed when detectors are
moved from the Fraunhofer ($r\gg L_{{\rm F}}$) to Fresnel ($r\ll
L_{{\rm F}}$) regions.

To conclude, we have studied electron propagation through a random
array of scatterers characterized by parameter $\eta =\tau
_{\mathrm{tr}}/\tau _{q}$--the ratio of the transport, $1/\tau
_{\mathrm{tr}}$, and ``quantum'', $1/\tau _{q}$, elastic
scattering rates (associated with the transport, $\sigma
_{\mathrm{tr}}$, and total, $\sigma _{\mathrm{tot}}$, scattering
cross-sections, respectively). For a given type of a disorder, the
parameter $\eta $ describes the relative strength of backward and
forward scattering. We have considered the case of strong
mesoscopic scatterers (e.g., quantum antidots) of a typical size
$a$ greater than the electron de Broglie wavelength $\lambda $.
For a wide class of scatterer's shapes, namely for convex ones, we
have shown that $\eta $ does not depend on the scatterer's shape
and size. In particular, for specular scattering, $\eta $ is a
universal constant determined only by the dimensionality ($D$) of
the system: $\eta =2$ for $D=3$ and $\eta =3/2$ for $D=2$.

We acknowledge stimulating discussions with V. E. Kravtsov, A. Yu. Kuntsevich,
S. P. Obukhov, and V. M. Pudalov.

  This work
was supported by RFBR Grants (No. 03-02-17285 and 06-02-16744) and
``Nanostructures" program of Russian Academy of Sciences (V. I.
Yu.) and NSF Grant No. DMR-0308377 (D. L. M.). We acknowledge the
hospitality of the Abdus Salam International Center for
Theoretical Physics (ICTP) where part of this work was done.


\begin{thebibliography}{*}

\bibitem{cuprates}  \O. Fischer, M. Kugler, I. Maggio-Aprile, C. Berthod, and C.
Renner,
Rev. Mod. Phys. {\bf 79}, 353 (2007).
\bibitem{mang} M. F{\"a}th et al., Science {\bf 285}, 1540 (1999).
\bibitem{karlsruhe} See, e.g., D. G. Polyakov, F. Evers. A. D. Mirlin, and P. W{\"o}lfle,
\prb {\bf 64}, 205306 (2001) and references therein.
\bibitem{antidot} C. Nachtwei, Z. H. Liu, G. L{\"u}tjering, R. R. Gerhadts, D.
Weiss, K. von Klitzing, and K. Eberl, \prb {\bf 57}, 9937 (1998).
\bibitem{ando}T. Ando, A. B. Fowler, F. Stern, Rev. Mod. Phys. {\bf 54}, 437
(1982).
\bibitem{Light}  H.\ C. van de Hulst, \emph{Light Scattering by Small
Particles}, (Dover Publications, New York, 1981), Section 8.

\bibitem{Sphere}  L. D. Landau and E. M. Lifshitz,{\em Quantum Mechanics}, 4th
edition, (Pergamon, 1977).

\bibitem{chaos} C. P. Dettmann, pp 315-365, in {\em Hard ball systems
 and the Lorentz gas}
edited by D. Szasz,
Encyclopaedia of Mathematical Sciences (Springer, 2000), v. 101, p. 315.

\bibitem{polyakov} D. G. Polyakov, F. Evers, and I. V. Gornyi, \prb {\bf 65},
125326 (2002).
\end{thebibliography}
\end{document}